\documentclass[fleqn,10pt]{wlscirep}
\usepackage[utf8]{inputenc}
\usepackage[T1]{fontenc}
\title{A network-based microfoundation of Granovetter's threshold model for
  social tipping}

\author[1,*]{Marc Wiedermann}
\author[2]{E. Keith Smith}
\author[1]{Jobst Heitzig}
\author[3,4]{Jonathan F. Donges}
\affil[1]{Complexity Science, Potsdam Institute for Climate Impact Research,
  Member of the Leibniz Association, P.O. Box 60 12 03, 14412 Potsdam, Germany}
\affil[2]{GESIS --- Leibniz Institute for the Social
  Sciences, Member of the Leibniz Association, Unter Sachsenhausen 6-8, 50667
  Cologne, Germany}
\affil[3]{Earth System Analysis, Potsdam Institute for Climate Impact Research,
  Member of the Leibniz Association, P.O. Box 60 12 03, 14412 Potsdam, Germany}
\affil[4]{Stockholm Resilience
  Centre, Stockholm University, Kr\"aftriket 2B, 114 19 Stockholm, Sweden}

\affil[*]{marcwie@pik-potsdam.de}

\begin{abstract}
Social tipping, where minorities trigger larger populations to engage in
collective action, has been suggested as one key aspect in addressing
contemporary global challenges. Here, we refine Granovetter's widely
acknowledged theoretical threshold model of collective behavior as a numerical
modelling tool for understanding social tipping processes and resolve issues
that so far have hindered such applications. Based on real-world observations
and social movement theory, we group the population into certain or potential
actors, such that -- in contrast to its original formulation -- the model
predicts non-trivial final shares of acting individuals. Then, we use a network
cascade model to explain and analytically derive that previously hypothesized
broad threshold distributions emerge if individuals become active via social
interaction. Thus, through intuitive parameters and low dimensionality our
refined model is adaptable to explain the likelihood of engaging in collective
behavior where social tipping like processes emerge as saddle-node bifurcations
and hysteresis.
\end{abstract}
\begin{document}

\flushbottom
\maketitle

\thispagestyle{empty}

\section*{Introduction}
Studies of collective behavior or action, such as protest demonstrations,
responses to disasters or even
revolution~\cite{SnowSocialMovementsCollective1995}, fosters an understanding
of the formation and logic of the
\textit{crowd}~\cite{ParkCrowdPublic1904,BlumerCollectiveBehavior1939,LoflandProtestStudiesCollective1985,McPhailMythMaddingCrowd1991}.
Broadly, the study of collective behavior can be separated into either those of
{\em social movements} or {\em temporary gatherings}. Social movements are
usually more structured around specific, identified goals, have deeper social
connections between actors, are organized  (generally to defend or fight
against existing authorities) and persist over time (such as the civil rights
movements)~\cite{DianiConceptSocialMovement1992}. In contrast, gatherings (such
as riots, sudden protests, concerts, sporting events) are more spontaneous,
less organized, do not carry as deep of social connections between actors, and
can be quite ephemeral~\cite{SnowMappingTerrain2004,
  McPhailCrowdCollectiveBehavior2006}. 

Further, individual engagement in collective behaviors (such as changing
consumption behavior or adoption of new technologies) can be connected to
broader social processes, such as norms and expectations for
behavior~\cite{nyborg_social_2016}. Specifically, individuals strategically
control their actions in accordance with their norms in order to achieve their
goals and
objectives\cite{McPhailBlumerTheoryCollective1989,McPhailMythMaddingCrowd1991,LoflandProtestStudiesCollective1985}.
As such, norms and preferences structure an actor's likelihood to engage in
collective behaviors, as well as its form of participation within these groups.
Complex forms of collective behaviors (be it either a movement or a crowd) are
thus created through dynamic interactions of actors that share common goals and
objectives for a given social situation. For example, global climate change has
been frequently noted as one prominent contemporary social problem that could
trigger and might also be addressed through collective behaviour (such as the
emergent `Fridays for Future'~\cite{hagedorn_concerns_2019}
movement)~\cite{farmer_sensitive_2019, moser2007toward}. 

Empirical evidence for such complex contagion of interlinked individuals that
leads to collective action has been provided for both
online~\cite{monsted_evidence_2017, centola_spread_2010,
  karsai_marton_complex_2014} and offline~\cite{christakis_social_2013} social
networks. Additionally, complex contagion has been experimentally shown to
foster social tipping~\cite{centola_experimental_2018}, a process that has
gained increased attention in the recent past~\cite{milkoreit_defining_2018}
due to its potential for rapid societal changes with profound impacts on the
entire socio-ecological Earth System~\cite{bentley_social_2014,
  moser2007toward}. Complementing empirical studies, recent conceptual models
of complex contagion  incorporate the spreading of an action, behaviour or
trait through a complex
network~\cite{house_thomas_modelling_2011,guilbeault_complex_2018,vespignani_modelling_2012, melnik_multi-stage_2013, watts_influentials_2007}.
They often aggregate an individual's surrounding over
time~\cite{dodds_universal_2004,dodds_generalized_2005} or abstract
space~\cite{watts_simple_2002} to accumulate exposure to a considered trait
such that at a certain point the individual adopts that trait as well. Such
models have been applied successfully to study processes involved in the
spreading of opinions~\cite{ho?yst_social_2001,hegselmann_opinion_2005},
large-scale epidemics~\cite{vespignani_modelling_2012}, the adoption of
life-style choices~\cite{schleussner_clustered_2016} or the collective
behaviour of animal groups~\cite{aoki_simulation_1982,couzin_collective_2002}.
However, most such models of collective behaviour are often tailored to a
specific problem (both in the incorporated processes as well as the underlying
parameter set) and are thus often not transferable to different and novel
applications.

The \textit{Granovetter threshold model} is a comparatively early contribution
to this field, providing a core basis for subsequent and more contemporary
modeling attempts~\cite{granovetter_threshold_1978}.  This model aims to
explain the emergence of collective behaviors while noting that individual
norms and preferences are a crucial factor determining their development and
final outcome. In particular, when presented with a simple binary choice -- to
participate within a collective behavior or not -- each individual has a certain
activation threshold for participation. This measures the proportion of the
group that an individual would like to observe participating within the
collective behavior before they are willing to join themselves. The thresholds
emerge from the norms, preferences, goals and beliefs of each individual, e.g.,
representing a kind of trade-off between the costs and benefits of joining in
the behavior.
As such, the application of the threshold model, or variations thereof, is not
limited to simple crowd-like behaviors, such as protests and riots, but is
comparatively broad, encompassing collective behaviors e.g.,
voting~\cite{kaempfer_threshold_1993}, diffusion of
innovations~\cite{zeppini_thresholds_2014}, or
migration~\cite{hunter_migration_2005}, as well as classical social movements
such as the Monday Demonstrations in East Germany~\cite{lohmann_dynamics_1994}.
However, while by design the model is very flexible, it has mainly been used
for illustrative and theoretical purposes (including most applications outlined
above), but hardly applied as a numerical modeling tool. 

This paper identifies two major sets of issues that prevent broader application
of the Granovetter model and proposes extensions to resolve them.  First, under
often assumed threshold distributions (such as cut-off
Gaussians~\cite{granovetter_threshold_1978}) the model usually unrealistically
predicts either no-one or the entire population to eventually act. We resolve
this issue by drawing from real-world observations, social movement
and resource mobilization theories~\cite{mccarthyzald1977,jenkins1983resource}, as well as recent theoretical and numerical results regarding network spreading processes~\cite{singh_threshold-limited_2013, karsai_local_2016}
to extend the original model by classifying individuals as either certainly
active, certainly inactive, or contingently active. This causes the model to
display nontrivial equilibria in which a certain part of the contingent
individuals becomes active. Second, the emergence and shape of the threshold
distribution itself is often underexplained. Therefore, we utilize an established conceptual
network cascade model~\cite{watts_simple_2002} and show that a broad
(non-Gaussian) threshold distribution emerges from microscopic networked
interactions in which potentially active individuals join an action if a
sufficient number of their neighbors are also engaged. We thus specifically acknowledge empirically observed tendencies of individuals to make decisions with respect to their immediate social surrounding rather than considering the entire global population, i.e., the mean field~\cite{centola_spontaneous_2015, centola_experimental_2018, garrod_conversation_1994}.  By addressing both of the above
issues, we effectively separate (unique) individual preferences which determine
general tendencies towards or against an action from the embedding of each
individual into a larger social structure and corresponding exposure to
external influences. Both characteristics then co-determine whether the
individual ultimately joins into an action or not. 

The remainder of this work is organized as follows. 
We first introduce the formal specifics of the Granovetter threshold model and
discusses in detail its aforementioned conceptual limitations.
We then implement the proposed solutions and 
present
a refined
threshold model that only depends on parameters that are readily observable in
real-world systems. Additionally, we provide an analytical solution of the
refined model and analyse its potential for modeling social tipping.
Ultimately, 
we close with a discussion of the results and
an outlook to future work.

\section*{Granovetter's threshold model}\label{sec:granovetter}

The threshold model assigns each individual in a population of size $N$ a
threshold that defines the number of others that must participate in an action
before the considered individual does so,
too~\cite{granovetter_threshold_1978}. In its discrete-time formulation the
number of acting individuals at time $t+1$, $R(t+1)$, is hence directly derived
from the cumulative distribution function of thresholds in the population, $F$,
such that

\begin{align}
    R(t+1)=NF(R(t)).\label{eqn:orig_granovetter}
\end{align}
Note that the original exemplary application of the model was that of
individuals' participation in riots. Hence the choice of the variable $R$ for
the number of acting individuals.  An equilibrium number of acting individuals
$R^*$ is obtained by solving $R(t+1)=R(t)=NF(R(t))$ for $R(t)$ which is
equivalent to finding an intersection of the graph of $F$ with the diagonal
through $(0,0)$ and $(N, N)$, Fig.~\ref{fig:fig1_scheme}a. All equilibrium
points $R^*$ at which $F$ intersects the diagonal line from above are stable,
while all others are unstable~\cite{granovetter_threshold_1978}.

While the threshold model has been widely used within a broad
literature~\cite{strang1998diffusion, dimaggio1983iron, jenkins1983resource} it
has up to now been mainly used for illustrative purposes as a number of issues
hinder its application as numerical modeling tool:

\subsubsection*{Plausible distributions typically predict no one or the entire population to act}\label{sec:granovetter_A}
As thresholds are hard to estimate, one typically assumes Gaussian threshold
distributions~\cite{granovetter_threshold_1978} cut off at the extreme values
$0$ and $N$. However, assuming a mean threshold $\mu$ of reasonable size and a
moderate standard deviation $\sigma$ implies that there are only few
individuals with low or high thresholds and many with medium thresholds close
to $\mu$. Hence, under the typical assumption of a low number of
\textit{instigators}~\cite{granovetter_threshold_1978} the model usually
predicts zero eventually acting individuals, Fig.~\ref{fig:fig1_scheme}a. Only
if a sufficiently large $\sigma$ is chosen more individuals than the
instigators become active. However, the choice of a large $\sigma$ causes the
distribution to become rather flat instead of bell-shaped. For example, for a
population size of $N=100$ and an average threshold of $\mu=25$, a standard
deviation of $\sigma=12.2$ is required so that a single instigator can cause
the rest of the population to become active~\cite{granovetter_threshold_1978}. 

In addition, if no individual has a threshold larger than $100\%$, the
threshold model generally has a second typically stable fixed point at $R^*=N$
implying that the entire population has the potential to become active if only
enough others do so, too, Fig.~\ref{fig:fig1_scheme}a. In reality, an
individual may never engage an action regardless of how many others have
already joined as personal preferences, norms or attitudes can restrict
behaviours~\cite{nyborg_social_2016}. In its basic setup, the Granovetter model
can only account for this by either assigning the concerned individuals a
threshold of 100\% or by selecting the population such that only those
individuals that are generally in favour of a certain action are
considered~\cite{granovetter_threshold_1978}. The first approach, however,
implies that everyone would generally be willing to act if only enough other
individuals become active before. The second approach requires updating the
population and, hence, its size, whenever the norms and attitudes of an
individual change. What both approaches have in common is that they imply a
constant change of the threshold distribution whenever individuals alter their
preferences or attitudes.

We therefore propose below a framework that refines the threshold model and accounts for
the above issues by grouping
individuals according to basic preferences that determine whether they
certainly, contingently or never act. This circumvents the existence of trivial
solutions and we show that this approach does not require a constant updating
of the threshold distribution as a response to changing group memberships. 

\subsubsection*{The threshold distribution can not be observed, but emerges from microscopic factors}\label{sec:granovetter_C}
Broadly, two complementary aspects shape whether an individual engages in an
action or not. On one hand there are {\em individual} factors (such as
background characteristics, social class, education or
occupation~\cite{CorningMyers2002,paulsen1991education}), that determine the
acceptance of or inclination towards an action. On the other hand there are
{\em group} factors, characteristics resulting from one's embedding in a larger
social network (such as social position, influence, or peer
pressure~\cite{LimSocialNetworks2008}). Both traits and processes ultimately
co-determine the macroscopic threshold that is exposed to the observer and we
call these thresholds of the original Granovetter model {\em emergent
  thresholds} from here on. However, quantifying the emergent thresholds on the
individual basis is difficult, if not impossible, to achieve without any prior
knowledge or assumptions on the aforementioned microscopic characteristics and interactions.
In addition, even properly justifying a certain shape of the emergent threshold
distribution is a difficult task as it remains unclear to which extent
different shapes follow from a certain composition of individual traits.

Notably, in analogy to the concept of emergent thresholds there should still
exist on the micro-level a share (or number) of others that join into an action
before an individual does so, too. One commonly accepted definition of such a
quantity is that of a {\em threshold fraction}~\cite{watts_simple_2002} that is
not assessed with respect to the entire population, but with regard to the
relevant social ties of a considered
individual~\cite{granovetter_threshold_1978, schelling_hockey_1973}. The
specific importance of one's egocentric social network for decision making has
recently been shown in empirical studies where individuals generally did not
aim for consensus or convergence in the global population, but rather on the
microscopic or group-level~\cite{centola_spontaneous_2015,
  centola_experimental_2018}. Additionally, it was observed that individuals
tend to coordinate with (at least subsets of) an entire group rather a single
partner~\cite{garrod_conversation_1994}. This renders the use of a
per-individual threshold fraction particularly useful as it determines the
share of others within a group that must make a certain decision before the
considered individual does so, too. In our specific case this threshold
fraction is considered a fundamental trait of each individual, regardless of
whether their preferences and norms favour or hinder a certain action. As such
it disentangles social processes from non-social factors, such as individual
preferences and norms. In contrast to the emergent thresholds, these threshold
fractions may not necessarily be widespread.  Rather, they might be  assumed to
have a narrow distribution or correspond to fixed, intuitive points, e.g. 50\%
(majority rule)~\cite{SchellingDynamicmodelssegregation1971}. Note that in
contrast to the emergent thresholds, that measure {\em absolute} numbers in a
global population, the threshold fraction measures the {\em relative} number of
others in one's egocentric social network that must make a decision before a
considered individual does so, too. It thereby specifically accounts for
heterogeneities in the number of each individual's neighbors, i.e., the
so-called social network's degree distribution~\cite{newman_networks:_2010}.

Below we present a microscopic threshold model
based on a previous study of cascading dynamics~\cite{watts_simple_2002} where
individual preferences are assigned to each member of the population that then
join into an action based on their threshold fractions applied to the
neighborhood in their social network. We then show that such microscopic
processes in fact yield an often postulated broad (but not normal-shaped)
emergent threshold distribution.

\section*{Results}

\subsection*{Refinement of the Model}\label{sec:refinement}
We start by addressing the first two issues identified above, namely that for
the usually chosen distributions the original model predicts either no-one or
the entire population to become active. As discussed above, one way to
circumvent these issues is to assign certain individuals either a threshold of
0\% or $\geq$100\% such that some individuals \textit{certainly} become active
and others never become active~\cite{granovetter_threshold_1978}. This approach
requires a constant updating of the threshold distribution and may be
impracticable for many cases. Recent studies investigated the effects of either such certainly active {\em initiators}~\cite{singh_threshold-limited_2013} or never active {\em immune} individuals~\cite{karsai_local_2016} on the adoption of certain traits or behaviours via spreading dynamics on social networks. In alignment with social movement theory~\cite{mccarthyzald1977, jenkins1983resource} we combine these two notions and suggest to divide the
population of size $N$ into three groups, namely: $A<N$ {\em certainly acting}
individuals~\cite{singh_threshold-limited_2013}, $C<N-A$ {\em contingent} individuals and the remaining $N-C-A$
{\em certainly inactive} individuals~\cite{karsai_local_2016}. The certainly acting and contingent
individuals form the group of $P=A+C$ \textit{potentially acting} individuals.
In a social movement and resource mobilization context, our three groups can
for example be seen as representing adherents, potential supporters and those
in opposition~\cite{mccarthyzald1977,jenkins1983resource}. 

\begin{figure}[t]
\centering
\includegraphics[width=\textwidth]{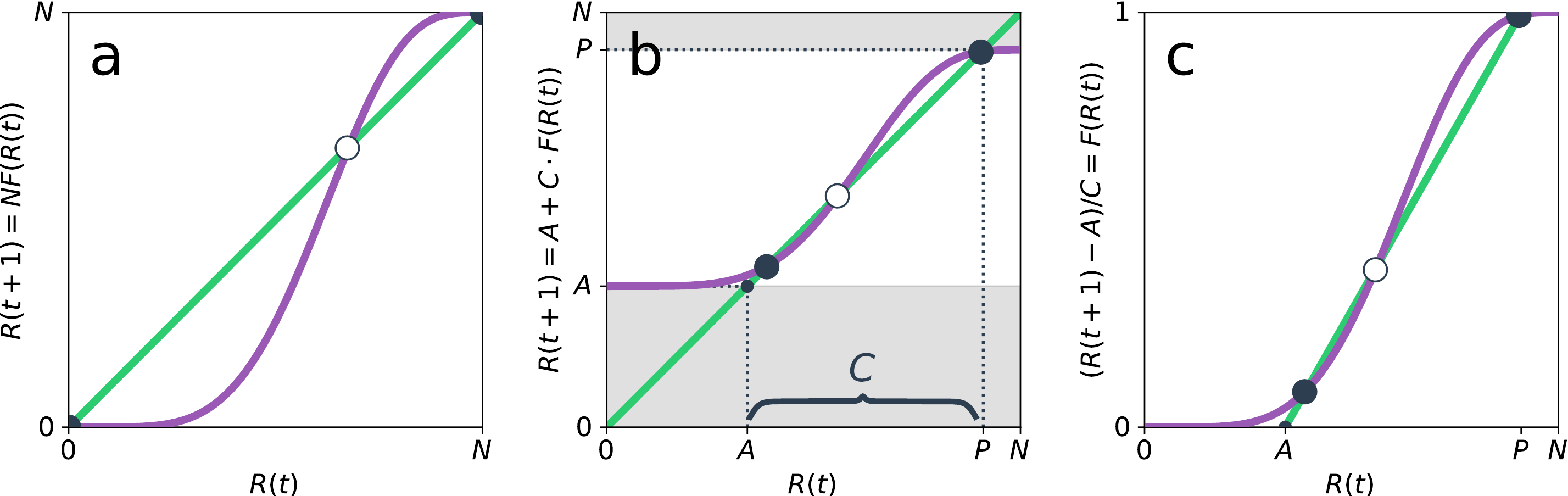}
\caption{ Extension of Granovetter's (graphic) model with $P$ potentially and $A$ certainly
  acting individuals. (a) The original model that computes the number of acting
  individuals
  $R(t+1)$ from the cumulative distribution function of thresholds
  $F$. The purple line indicates a typical normal-like choice for this
  distribution. The $45^\circ$-line (green) intersects $F$ at the 
  stable (black) and unstable (white) equilibrium points $R^*$. As for many realistic
  choices of $F$, only $R^*=0$ and
  $R^*=N$ are stable. (b) Introducing
  $A$ certainly and $P$ potentially acting individuals, such that the $C=P-A$
  contingent individuals have the same threshold distribution $F$ as the entire
  population $N$. Here, the equilibria move to the interval $R^*\in[A,P]$ and are not
  necessarily located at exactly $R^*=A$ and $R^*=P$. Hence, the $A$ certainly
  acting individuals trigger some contingent individuals to act, too. (c) Rescaling
  $R(t+1)$ to the unit interval shows that equilibria can be computed by
  shifting the diagonal line from crossing $(0,0)$ and $(N,N)$ (as in (a)) to
  crossing $(A,0)$ and $(P,1)$ and using the same threshold distribution
  $F$ as in
(a).}
\label{fig:fig1_scheme}
\end{figure}

If we have no reason to assume that the threshold distribution is different in
the three groups, the original recursive formula Eq.~\eqref{eqn:orig_granovetter}
is then replaced by

\begin{align}
    R(t+1) = A + C\cdot F(R(t)).\label{eqn:adapted_granovetter}
\end{align}
The equilibria of the thus refined model are again obtained by computing the
intersection of the r.h.s.\ of Eq.~\eqref{eqn:adapted_granovetter} with the
diagonal through $(0,0)$ and $(N,N)$, Fig.~\ref{fig:fig1_scheme}b. It is
apparent that if $A>0$ and $P<N$ (note again that $P=A+C$), we get nontrivial
equilibrium numbers of acting individuals $R^*\in[A, P]$. Conveniently, as $A$
or $P$ (and $C$) change, the new equilibria can be found without re-estimating
the threshold distribution.

In order to also avoid having to redraw $F$ in Fig.~\ref{fig:fig1_scheme}b
whenever there is a variation in $A$ or $C$, it is beneficial to rescale the
ordinate to the unit interval, Fig.~\ref{fig:fig1_scheme}c. This allows us to
find the equilibria for all possible combinations of $A$ and $P$ in the same
diagram, by drawing $F$ only once and just adjusting the diagonal to meet the
points $(A, 0)$ and $(P, 1)$.

Our adjusted approach makes the application of the threshold model as an actual
modeling framework more practical as it (i) produces nontrivial fixed points
$R^*$, (ii) requires the threshold distribution to be only estimated once for
the entire population or a representative sample thereof, and (iii) relies on
only two intuitive parameters, the size of the certainly ($A$) and potentially
acting population ($P$). Recall that $A$ directly relates to an immediate
action or behaviour, while $P$ denotes the general acceptance of or attitude
towards that action. 

\subsection*{Estimation of the emergent threshold distribution}\label{sec:results_estimation}
Having refined the threshold model to properly allow for the computation of
non-trivial fixed points, we shift our focus to the second  issue that relates
to the threshold distribution itself. It has been established 
above
that the emergent thresholds follow from microscopic characteristics of each
individual as well as its embedding in a social context. Specifically for the
latter it will turn out that the share of others, i.e., the threshold fraction, that must join into an
action before a contingent individual does so too need not be widely
distributed or even heterogeneous at all across the population in order to
produce a widespread distribution for the emergent threshold.

We now study how such characteristics and interactions on the micro-level
determine one's emergent threshold by using a simulation model of social
contagion that has been studied in the past to model binary decisions with
externalities and resulting cascading dynamics~\cite{watts_simple_2002}. We represent each individual in the
population by a node in a complex network and draw links between nodes to
indicate their embedding in a social group of others (see
Methods section below
for details). This relates directly to the idea of
a \textit{sociomatrix} that accounts for the stronger influence that
individuals to which one forms a social bond have on one's
behaviour~\cite{granovetter_threshold_1978}. In addition to the original formulation of this network cascade model~\cite{watts_simple_2002} and in agreement with the
consideration put forward 
above
we assume that $P$
randomly distributed nodes form the potentially active population. Being
potentially active subsumes all norms, preferences and attitudes that cause an
individual to show acceptance for a considered type of behaviour.  Among the
$P$ potentially active nodes we assume that $A\leq P$ randomly distributed
nodes are certainly active. In each time step each of the remaining $C=P-A$
contingent nodes $i$ becomes active if more than a share $\varrho\in[0,1]$ of
its immediate neighbors is already active. We hence denote $\varrho$ the {\em threshold fraction} of an individual. The resulting number or active nodes at time
$t$ is again denoted as $R(t)$. Setting a common value of $\varrho$ represents
the most narrow distribution of actual threshold fractions that determine
whether one joins into action given that one generally supports that action at
all.

\begin{figure}[t!]
\centering
\includegraphics[width=.7\linewidth]{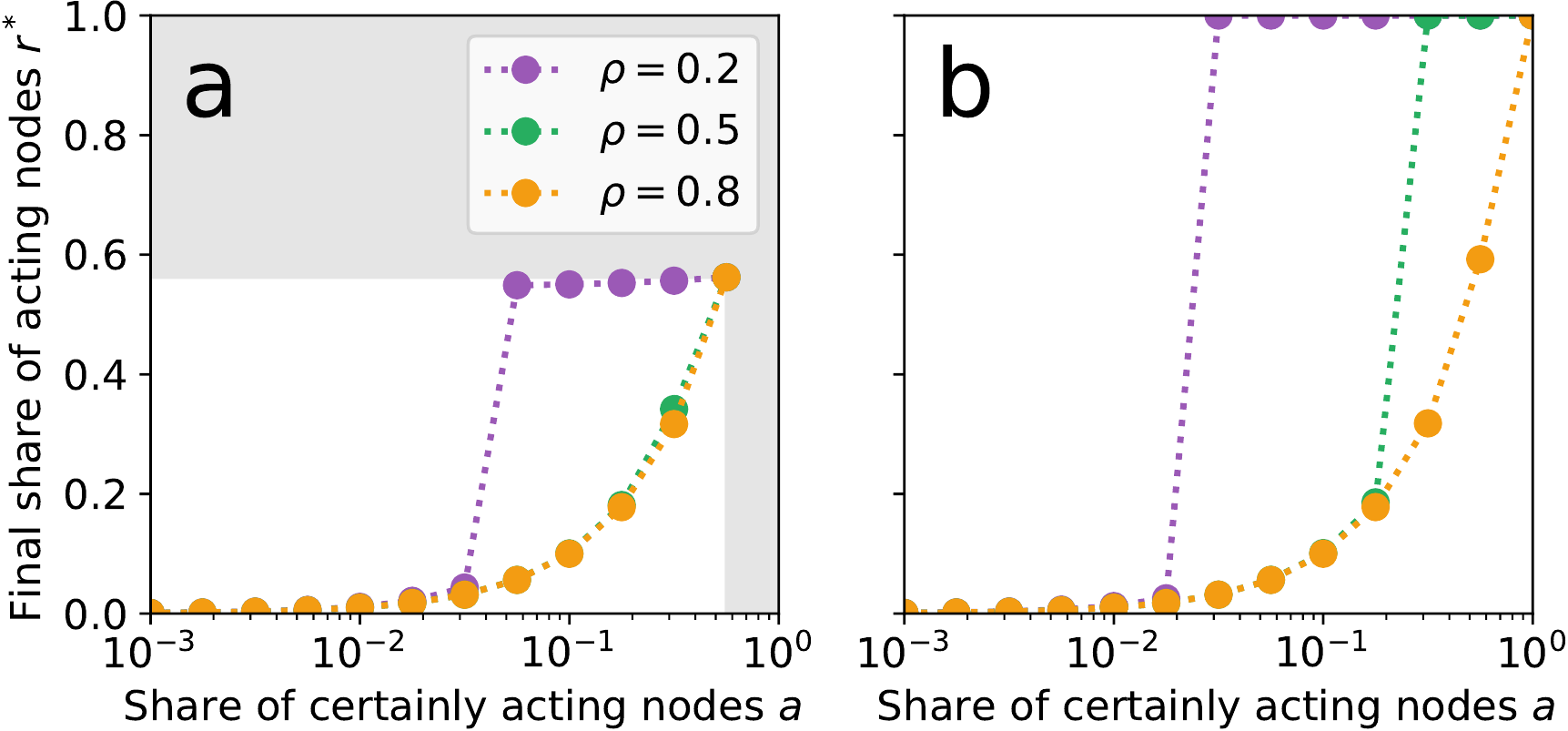}
\caption{The final share of acting nodes $r^*$ in the microscopic network
  simulation for given shares of certainly acting nodes $a$. (a) With only
  around half the population being potentially active (i.e, $p=P/N\approx
  0.56$) only a low threshold fraction ($\varrho=0.2$, purple) causes large shares
  of the contingent nodes to act. Grey areas indicate values of $r^*$ and $a$
  that would exceed $p$. (b) If every node in the network is potentially active
  ($p=1$), also an intermediate threshold fraction ($\varrho=0.5$, green) suffices
  to cause the entire population to act. In comparison with (a) one also
  observes that the transition observed for $\varrho=0.2$ occurs already for
  smaller choices of $a$. For a large threshold fraction ($\varrho=0.8$, yellow)
  no abrupt transition appears such that $r^*\approx a$ for all considered
  choices of $a$ and $p$.}
\label{fig:final_active}
\end{figure}

We simulate cascades of nodes becoming active for two different shares of
potentially active nodes $p=P/N=0.56$ (Fig.~\ref{fig:final_active}a) and $p=1$
(Fig.~\ref{fig:final_active}b), as well as for different threshold fractions
$\varrho\in\{0.2, 0.5, 0.8\}$. Fig.~\ref{fig:final_active} shows the final
share of acting nodes $r^*=R^*/N$ after the cascade stops for increasing shares
of certainly acting nodes $a=A/N\leq p$. For $p=0.56$ (i.e., a low share of
potentially acting nodes) only small threshold fractions ($\varrho=0.2$) allow for
a large-scale cascade such that $r^*\rightarrow p$ for values of $a\gtrapprox
0.05$ (Fig.~\ref{fig:final_active}a). In contrast, for values of $a\lessapprox
0.05$ no cascade is observed and, hence, $r^*\approx a$. Larger
threshold fractions (i.e., $\varrho=0.5$ or $\varrho=0.8$) hinder the emergence of
a cascade such that $r^*\approx a$ for all choices of $a$
(Fig.~\ref{fig:final_active}a).  For $p= 1$, cascades are also observed at a
larger threshold fraction of $\varrho=0.5$ but are still suppressed for
$\varrho=0.8$ (Fig.~\ref{fig:final_active}b). Furthermore, the required share
of certainly acting nodes $a$ at which the system \textit{tips} from a state
with no cascades to a state with a global cascade decreases slightly with
increasing $p$ (compare Fig.~\ref{fig:final_active}a and
Fig.~\ref{fig:final_active}b). Note that specifically the role of the remaining
$N-P$ certainly inactive nodes has been studied under the term {\em immune}
nodes in an earlier study of spreading dynamics on
networks~\cite{karsai_local_2016}. However, in contrast to our results presented above the underlying model in this previous work~\cite{karsai_local_2016} assumed the share of certainly active nodes $a$ to increase over time at a constant rate, thus yielding convergence to a globally stable fixed point $r^*=p$ for all initial choices of $a$. Hence, the major purpose of the {\em immune} nodes in this earlier work was to moderate the rate of convergence to that global fixed point.

\begin{figure}[t!]
\centering
\includegraphics[width=.5\linewidth]{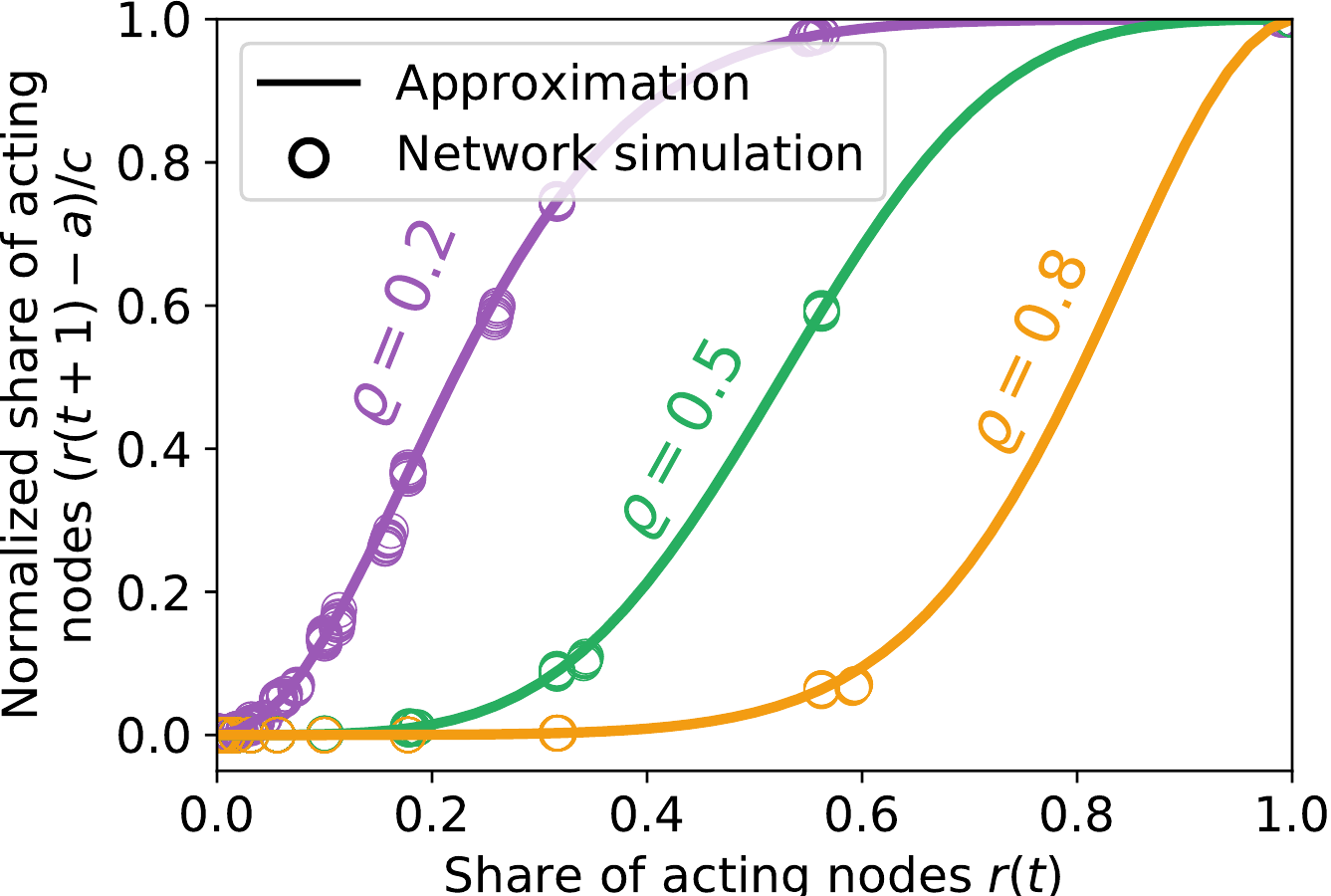}
\caption{Emergent threshold distribution measured from the microscopic network
  simulations
  and the analytical approximation. For the
  network simulations only those points where the system is close to
  equilibrium, i.e. $t\in \{0, t_{max}-1\}$, are shown. For all shown choices of threshold fractions $\varrho$, the approximation
  matches well with the network simulations.}
\label{fig:analytical_thresholds_with_data}
\end{figure}

To estimate an emergent threshold distribution as required for the Granovetter-type threshold
model we now plot $r(t)=R(t)/N$ against $(r(t+1) - a)/c$ (with $c=C/N=(P-A)/N$)
from the network simulations. Fig.~\ref{fig:analytical_thresholds_with_data}
shows the results if the network cascade is close to equilibrium, i.e., for
$t=0$ or $t=t_{max}-1$, where $t_{max}$ is the time at which the cascade stops.
We observe the formerly postulated broad distribution of emergent thresholds as
a result of the microscopic interactions at narrowly distributed
threshold fractions $\varrho\in\{0.2, 0.5, 0.8\}$ given a generally positive ($P$
nodes) or negative ($N-P$ nodes) attitude towards the considered behaviour.
This implies that individuals with a high emergent threshold may not
necessarily be more reluctant to join into an action, it could simply mean that
they are located at a more peripheral position in the network.  

By approximating the number of active, $a_i$, and inactive neighbors, $b_i$, of
a node $i$ as coming from a common multinomial distribution that only depends
on the number of neighbors $k_i=a_i+b_i$ and the overall share of active nodes
$r(t)$,  we derive an analytical approximation of the emergent threshold
distribution $F$ (note that for brevity we omit the dependence of $r(t)$ on
$t$) as 

\begin{align}
    F(r)&= 1-\exp(-K)\sum_{b_i=0}^{\infty}\frac{(K-Kr)^{b_i}}{b_i!}
\sum_{a_i=0}^{ \left\lfloor \frac{\varrho b_i}{1-\varrho}\right\rfloor
}
\frac{(Kr)^{a_i}}{a_i!}.\label{eqn:approximation}
\end{align}
Here, $K = \sum_i k_i / N$ denotes the average degree (i.e., number of
neighbors) of nodes in the network (see 
Methods section below
and
the Supplementary Information for a 
full derivation of Eq.~\eqref{eqn:approximation}). Note that the
second factor in Eq.~\eqref{eqn:approximation} can be further approximated by an
incomplete Gamma-Function. We find that (close to equilibrium)
Eq.~\eqref{eqn:approximation} aligns very well with the network simulations for small ($\varrho=0.2$), medium ($\varrho=0.5$) and large ($\varrho=0.8$) fractional thresholds (Fig.~\ref{fig:analytical_thresholds_with_data}) and thus complements previously proposed approximations that primarily held for small to medium values~\cite{singh_threshold-limited_2013}. For
the transient phase the approximation still estimates the emergent thresholds
well for small and large choices of $\varrho$ but decreases in quality for
intermediate values (see Supplementary Information). This is mainly caused by the clustering of active
and inactive nodes. An extension of the above approximation that accounts for
such factors, e.g., via pair approximations \cite{wiedermann_macroscopic_2015, gleeson_binary-state_2013} or moment generating functions~\cite{watts_simple_2002},
is beyond the scope of this work and remains as a subject for future research.
In summary, Eq.~\eqref{eqn:approximation} gives a good estimation of an emergent
macroscopic distribution that fulfills the initially postulated broad
shape~\cite{granovetter_threshold_1978} while emerging from a subsumed set of
preferences as well as a single common threshold fraction $\varrho$. In addition,
using a single distribution $F$ has the advantage of being independent of the
share of certainly and potentially acting nodes. As such it only needs to be
estimated once while changing preferences (i.e., varying $A$/$a$ and $P$/$p$)
are incorporated into shifting the diagonal line that is used to estimate the
fixed points (see again Fig.~\ref{fig:fig1_scheme}c).   

\subsection*{Comprehensive analysis and social tipping}
\begin{figure}[t!]
\centering
\includegraphics[width=.8\linewidth]{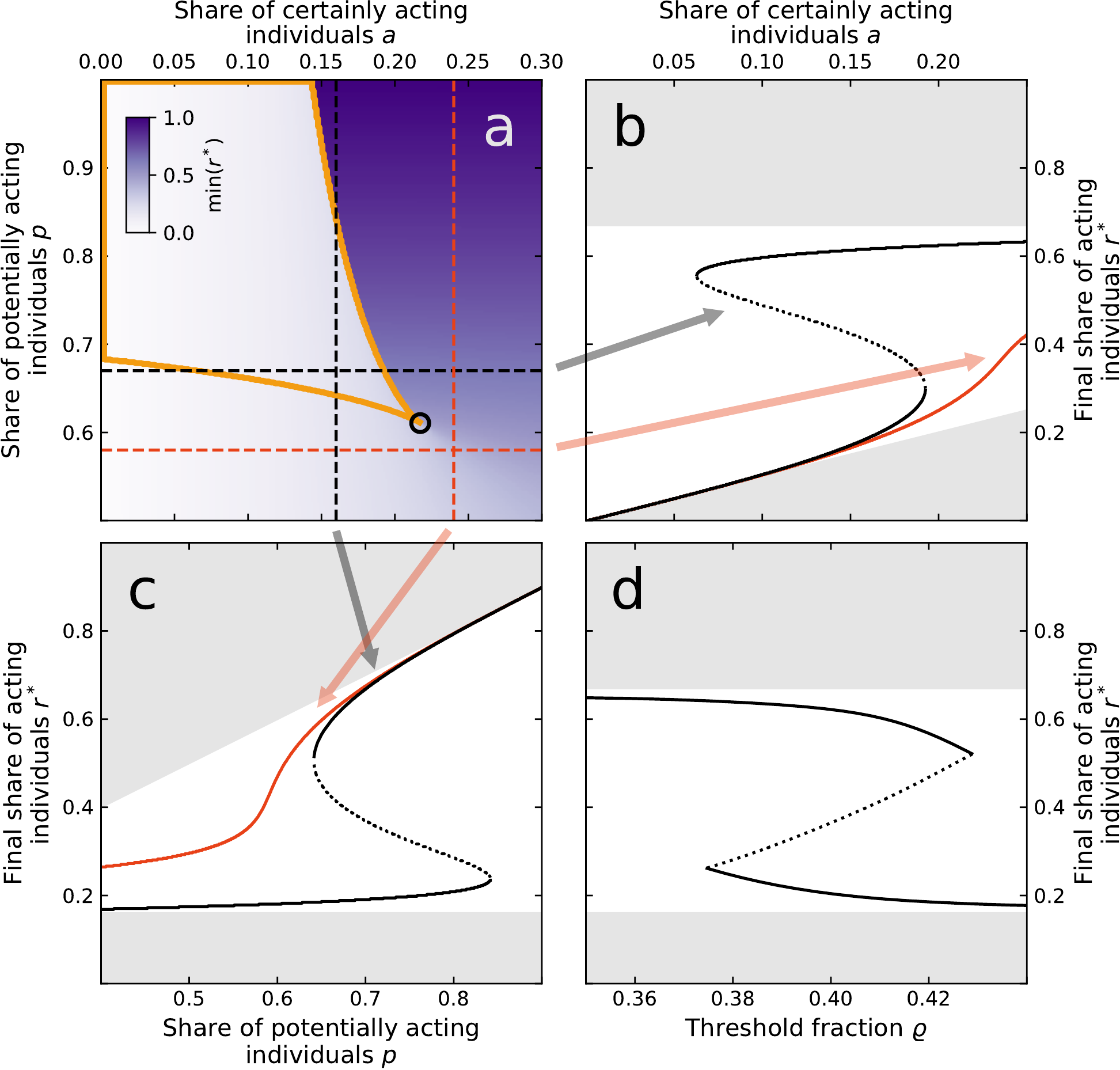}
\caption{
  Bifurcation analysis and hysteresis of the refined Granovetter model
    with an emergent threshold distribution as given by the analytical
    approximation. (a) Smallest stable fixed point $\min(r^*)$ for different
  shares of certainly acting $a$ and potentially acting individuals $p$. The
  black circle denotes a cusp-bifurcation. Black dashed horizontal/vertical
  lines correspond to the diagrams in (b)/(c) that show a saddle-node
  bifurcation. For (b),(c) and (d), solid (dotted) lines indicate stable
  (unstable) fixed points $r^*$. Grey shading indicates those areas where
  $r^*\notin [a,p]$ and that can thus not be reached. The yellow circled area
  in (a) indicates the bistable regime. Red lines in (a) correspond to values
  of $p$ and $a$ at which no bifurcation is observed and thus $r^*$ varies
  smoothly in (b)/(c). (d) shows the bifurcation diagram in the threshold fraction
  $\varrho$. Fixed parameters are: $a=0.16$ for (c) ($a=0.24$ for the red line)
  and (d), $p=0.67$ for (b) ($p=0.58$ for the red line) and (d), and
  $\varrho=0.4$ for (a), (b) and (c).}
\label{fig:bifurcation}
\end{figure}

From the approximate emergent threshold distribution $F$ in
Eq.~\eqref{eqn:approximation} we estimate the fixed points $r^*$ of the refined
threshold model for different choices of $a$, $p$ (or $c=p-a$), and $\varrho$
by solving $(r-a)/c=F(r)$ (i.e., intersecting the diagonal line with $F$). We
either identify two stable and one unstable fixed points, or one globally
stable fixed point $r^*$. Fig.~\ref{fig:bifurcation}a shows the value of the
smallest stable fixed point $\min(r^*)$. We find a sharp increase in its value
for certain values of $0.15\lessapprox a\lessapprox 0.22$ and $p\gtrapprox 0.5$
hinting at a saddle-node bifurcation. Fig.~\ref{fig:bifurcation}b,c show that
saddle-node bifurcation at varying values of $a$ and $p$, respectively. As the
saddle-node bifurcation, and correspondingly also hysteresis, emerges in both
parameters, the model consequently displays a cusp-bifurcation as well (see
black circle in Fig.~\ref{fig:bifurcation}a). For fixed values of $a$ or $p$
below the cusp-point the final share of acting individuals $r^*$ thus varies
only smoothly with the respective other free parameter (red lines in
Fig.~\ref{fig:bifurcation}b,c). In contrast, fixing either $a$ or $p$ to values
above the cusp-point can cause the system to rapidly shift from a stable state
with low $r^*$ to a stable state with high $r^*$ (and vice versa) as the
corresponding bifurcation point in the remaining free parameter is crossed
(black lines in Fig.~\ref{fig:bifurcation}b,c). Notably, the model shows
hysteresis also within a band of possible threshold fractions,
Fig.~\ref{fig:bifurcation}d). 

In summary, our model conceptually shows what has formerly been termed
\textit{social tipping}, i.e., a process where for a given population a small
change in the size of a dedicated minority can have a large
effect~\cite{centola_experimental_2018,pruitt_jonathan_n._social_2018,bentley_social_2014}.
In our specific case, for a given value of $a$ or $p$ a small change in the
respective other parameter suffices to largely increase (or decrease) the share
of finally acting individuals $r^*$. Complementing recent theoretical and numerical studies of spreading processes on networks that either varied the size of the initiating minority~\cite{singh_threshold-limited_2013} or the so-called immune group of inactive nodes~\cite{karsai_local_2016} our model shows a bistable regime that is necessary for the emergence of hysteresis. This implies that once the system has tipped it sustains its state of high (low) shares of acting individuals $r^*$ even if $a$
or $p$ were to be reduced (increased) again. By incorporating both, initiating and immune groups, our model additionally gives rise to a previously undetected cusp-bifurcation as well. 

Remarkably, the critical size of the dedicated minority at which the system undergoes a fold-bifurcation (Fig.~\ref{fig:bifurcation}a,b) has recently been empirically estimated to lie in the range $0.21\lessapprox a\lessapprox 0.25$ which is consistent with the results of our  model~\cite{centola_experimental_2018}. Moreover, critical minority group sizes of around 20 percent have also been discussed with respect to the Pareto-Principle~\cite{pareto_manual_1971} which has recently been reframed as {\em the law of the vital few} to discuss matters of sustainability transformations and social tipping~\cite{schellnhuber_why_2016}. 

\section*{Discussion}
We have proposed a refined version of the original Granovetter threshold
model~\cite{granovetter_threshold_1978} that addresses a set of issues that,
so far, have hindered its application as a conceptual modeling tool.
Specifically, we propose to divide the considered population of size $N$ into
three classes (certainly, potentially, and certainly not acting individuals) of
different sizes $A\leq P$, $P\leq N$, and $N-P$. In addition, we propose a
threshold distribution that emerges from microscopic interactions between
individuals on a social network. This distribution solely depends on the
average connectivity $K$ of individuals and a common threshold fraction $\varrho$
to join into an action given that their individual preferences and attitudes
are already favourable with respect to that action. The four parameters of our
refined model are of intuitive nature and allow for a systematic evaluation of
its dynamics in terms of a bifurcation analysis (except for $K$ which only
needs to be chosen sufficiently larger than zero, i.e., $K\gg 0$, see
Supplementary Information for
details). As in the original threshold model, an estimation of the fixed points
can be obtained by (graphically) intersecting the diagonal line defined by $a$
and $p$ with the emergent threshold distribution $F$. The three crucial
parameters $a$, $p$, and $\varrho$ all cause a saddle-node bifurcation which is
a prototypical mechanism behind tipping points in many other systems, such as
in ecology~\cite{beisner_alternative_2003,dai_generic_2012} or the climate
system~\cite{lenton_climate_2012, thompson_predicting_2011}, as well. It thus
makes the model a promising tool to study the emerging field of social
tipping~\cite{pruitt_jonathan_n._social_2018,centola_experimental_2018,bentley_social_2014}
where \textit{little things can make a big
  difference}~\cite{gladwell2006tipping} and minority groups can trigger large
shares of a population to engage in collective action. 

Our revised model describes multiple forms of collective behaviors, including
social movements and crowd-like behaviors. For both such behaviors, norms are
directly called upon to structure individual likelihood to engage in actions
while also observing the actions of others around them. Importantly, there are
differences in the speed of the process. For crowds the observation of social
members is made relatively quickly, as are the decisions to participate in the
actions. In contrast, these processes can be much slower for social movements.
For both cases, however, we identify three time scales that are underlying our
refined threshold model. We assume that the microscopic threshold fractions change at
the slowest time scale (usually years to decades), as these are attributed to
the unique identity of an individual (which may be less prone to sudden
external shocks). In contrast, the classification into {\em certainly} or {\em
  contingently} active individuals varies on intermediate time scales (months
to years) as changes in the environment (such as financial shocks or the
exposition to increasing extreme weather events) are beyond an individual's own
agency and can trigger sudden changes in attitudes~\cite{ricke_natural_2014}.
The social dynamics modelled here, i.e., the observation of others and the
joining into an action, are happening on the fastest time scale (days to
months) as frequent social interactions are common among members of any given
society.

Most parameters of the refined model may be readily measurable in a variety of
applications. Attitudes that determine $p$ could be estimated from surveys or
existing panel data. The share of certainly acting individuals $a$ could be
given by those in the population that inevitably need to act, e.g., migrate as
a consequence of climate change impacts~\cite{black_climate_2011,
  mcleman_migration_2006}. For the average degree $K$ it may often suffice to
set it to a reasonable number, e.g., Dunbar's Number that suggests a cognitive
limit to the number of people with whom an individual can maintain a persistent
social relationship~\cite{dunbar_coevolution_1993} (see Supplementary
Information for details). The
threshold fraction $\varrho$ could then either remain as a free parameter of the
model or be set to fixed intuitive points such as 50\% (majority rule) or 20\%
(Pareto principle~\cite{pareto_manual_1971, schellnhuber_why_2016}). Furthermore, the model also allows for changes in its
parameters over time, such that $r^*$ can be estimated as a time-dependent
variable, possibly causing the system to tip back and forth between its two
possible stable states. In that sense the respective parameters can be
incorporated into the system's internal dynamics as slowly changing variables. 

Future work should concentrate on collecting data for the different parameters
and then consequently test and calibrate the model against historical test
cases. One specific challenge that lies within such an endeavor is the
estimation of appropriate (relative) time scales at which the parameters and
the internal variables change. In addition, appropriate early-warning
indicators~\cite{scheffer_early-warning_2009,jiang_predicting_2018,
  thompson_predicting_2011} should be applied to study the existence of
precursory signals for the transgression of a social tipping point, i.e.,
bifurcation, in our model. Some of these indicators would require a further
extension of the model such that individuals may also spontaneously become
active with a low probability even if their threshold fraction is not transgressed
(or vice versa). We further acknowledge that up to now a proposal for an
emergent threshold distribution has only been derived analytically for the case
of an Erd\H{o}s-R\'enyi random network~\cite{erdos_evolution_1960}. While this
lays good groundwork, the threshold distribution should also be explored for
topologies (such as \textit{scale-free}~\cite{barabasi_emergence_1999} and
\textit{small-world} networks~\cite{watts_collective_1998}) that more closely
mimic those of real-world social systems. Hence, even though our proposed approximation of the emergent threshold distribution holds well if the system is well-mixed and close to a fixed point, more elaborate methods, e.g., pair approximations~\cite{gleeson_binary-state_2013} and moment generating function approaches~\cite{watts_simple_2002}, should be used to predict the model's dynamics for more general network topologies and during transient phases as well. Ultimately, the model should be
applied as a conceptual modeling tool, e.g., to make qualitative statements on
the possibility for social tipping with respect to issues of global change or
sustainability transformations~\cite{farmer_sensitive_2019,
  westley_tipping_2011, david_tabara_positive_2018} under different scenarios. 

\section*{Methods}

\subsection*{Network cascade model}\label{sec:matmet_model}
For the microscopic network simulation we consider an Erd\H{o}s-R\'enyi random
network~\cite{erdos_evolution_1960} with $N=100\,000$ nodes and a linking
probability of $\ell=9\cdot10^{-5}$ resulting in an average degree of $K=10$.
We vary the number of certainly acting nodes $A$ logarithmically between $1$
and $N$ and the number of potentially acting nodes logarithmically between $A$
and $N$. For each setting of $A$ and $P$ (and fixed values of the threshold fraction $\varrho$ as
given in Fig.~\ref{fig:final_active}) we create an ensemble of $n=100$ networks
and randomly assign $P$ out of the $N$ nodes as potentially active. Out of
those $P$ nodes we then randomly assign $A$ certainly acting nodes. The model
then runs in discrete time steps $t$. In each time step, every potentially
active, yet inactive, node $i$ becomes active if its share of active neighbors
exceeds the threshold fraction $\varrho$. All nodes update their status
synchronously at each time step. The simulation stops if the number of newly
activated nodes at time $t$ equals zero, i.e., if $R(t-1)=R(t)$. Note that our
model is based on previous works that implemented a simpler version of a
cascade model that did not account for a distinction in potentially active and
certainly inactive nodes~\cite{watts_simple_2002}. 

\subsection*{Approximation of the emergent threshold distribution}\label{sec:matmet_approximation}
The approximate emergent threshold distribution $F$ in
Eq.~\eqref{eqn:approximation} is derived by assuming that for each individual
$i$ the number of active $a_i$ and inactive neighbors $b_i$ are distributed
according to a common multinomial distribution, giving

\begin{align}
    F(R) &=\sum_{\substack{a_i>\varrho (a_i+b_i)\\a_i\leq R\\ b_i\geq0\\
        b_i\leq P'}} \binom{R}{a_i} \binom{P'}{b_i} \ell^{a_i} \ell^{b_i}
    (1-\ell)^{R-a_i} (1-\ell)^{P'-b_i}.\label{eqn:multinomial}
\end{align}
$P'=N-1-R$ denotes the number of inactive individuals that are not the
considered $i$, as one's own level of activity is not accounted for. $\ell$ is
the linking probability of the Erd\H{o}s-R\'enyi network.
Eq.~\eqref{eqn:approximation} follows from Eq.~\eqref{eqn:multinomial} by
setting $R=\lfloor rN\rfloor$, substituting the binomial distributions by two
Poisson distributions with expectation values $\lambda_a=Kr$ and
$\lambda_b=K-Kr$ and assuming that $N\gg K$. A step-by-step derivation of
Eq.~\eqref{eqn:approximation} is given in the Supplementary Information.


\section*{Acknowledgements}
This work was developed in the context of the COPAN
  collaboration at the Potsdam Institute for Climate Impact Research (PIK).
  M.W.\ and K.S.\ are supported by the Leibniz Association (project DOMINOES).
  J.F.D.\ is grateful for financial support by the Earth League’s EarthDoc
  program and the European Research Council advanced grant project ERA (Earth
  Resilience in the Anthropocene). The authors gratefully acknowledge the
  European Regional Development Fund (ERDF), the German Federal Ministry of
  Education and Research and the Land Brandenburg for providing resources on
  the high-performance computer system at PIK.

\section*{Author contributions statement}
All authors designed the study. M.W.\ performed the numerical simulations and
analysed the data. M.W.\ and J.H.\ derived the analytical\ approximation. M.W.\
and E.K.S.\ drafted the manuscript. All authors substantively revised the work.

\section*{Additional information}

\textbf{Competing interests} 
The authors declare no competing interests.

\clearpage

\clearpage

\clearpage

\end{document}